\documentclass{article}  

\usepackage{mathtools}
\usepackage{wrapfig}
\usepackage{pgf,tikz,pgfplots}
\usetikzlibrary{decorations.pathreplacing, calligraphy}
\usetikzlibrary{decorations.markings}
\usetikzlibrary{patterns}
\usetikzlibrary{calc}
\makeatletter 
\tikzset{ 
reuse path/.code={\pgfsyssoftpath@setcurrentpath{#1}} 
} 
\tikzset{even odd clip/.code={\pgfseteorule}, 
protect/.code={ 
\clip[overlay,even odd clip,reuse path=#1] 
(current bounding box.south west) rectangle (current bounding box.north east)
; 
}} 
\makeatother 
\usetikzlibrary{3d,perspective}

\usepackage{graphicx} 
\usepackage{amsmath,amssymb,amsbsy,amstext, amsthm, simplewick}
\usepackage{xcolor}
\usepackage{tcolorbox}
\usepackage{mdframed}
\mdfsetup{%
linecolor=white,
backgroundcolor=gray!20,
}
\usepackage{amsfonts}
\usepackage{jcappub}
\def\be{\begin{equation}}
\def\ee{\end{equation}}
\def\bea{\begin{eqnarray}}
\def\eea{\end{eqnarray}}

\def\wn{\widehat{n}}
\newcommand{\z}{\zeta}
\newcommand{\zb}{\overline{\zeta}}
\def\d{{\rm d}}

\def\lsim{\mathrel{\rlap{\lower4pt\hbox{\hskip0.5pt$\sim$}}
    \raise1pt\hbox{$<$}}}         
\def\gsim{\mathrel{\rlap{\lower4pt\hbox{\hskip0.5pt$\sim$}}
    \raise1pt\hbox{$>$}}}         

\makeatletter
\newlength{\apb@width}
\newcommand{\autoparbox}[2][c]{\settowidth{\apb@width}{#2}\parbox[#1]{\apb@width}{#2}}

\pgfplotsset{compat=1.18}
\begin{document}

\begin{titlepage}

\setcounter{page}{1} \baselineskip=15.5pt \thispagestyle{empty}

\bigskip\

\vspace{2cm}

\vspace{0.5cm}
\begin{center}

{\fontsize{20}{28}\bfseries The  PTA Hellings and Downs Correlation}

\vskip 0.3cm
{\fontsize{20}{28}\bfseries   Unmasked by Symmetries}
\end{center}

\vspace{0.5cm}

\begin{center}
{\fontsize{13}{30}\selectfont  A. Kehagias, $^{1}$ 
 and A. Riotto,$^{2}$}
\end{center}

\begin{center}

\vskip 8pt

\textsl{$^1$ Physics Division, National Technical University of Athens, Athens, 15780, Greece}
\vskip 7pt

\textsl{$^2$ Department of Theoretical Physics and Gravitational Wave Science Center (GWSC), \\
 Université de Genève, CH-1211 Geneva }
\vskip 7pt

\end{center}

\vspace{1.2cm}
\hrule \vspace{0.3cm}
{ \noindent 
\hskip 6.5cm
{\fontsize{12}{12} \bfseries Abstract} \\[0.1cm]
The Hellings and Downs correlation curve 
describes
 the correlation of  the timing residuals from  pairs of pulsars as a function of their angular separation on the sky and  is 
 a smoking-gun signature for the detection of an isotropic   stochastic background of gravitational waves. We show that it can be  easily obtained from 
realizing that Lorentz transformations are conformal transformations
on the celestial sphere and from the conformal properties of the two-point correlation of the timing residuals. This result allows several generalizations, e.g. the calculation of the three-point correlator of the time residuals and the  inclusion of  additional polarization modes (vector and/or scalar) arising  in alternative
theories of gravity.

\noindent}
\vspace{0.2cm}
\hrule

\vspace{0.6cm}

\end{titlepage}

 \tableofcontents

\newpage 
\baselineskip=18pt
\section{Introduction}
The recent Pulsar Timing Arrays (PTA) collaborations,   NANOGrav~\cite{NG15-SGWB,NG15-pulsars}, EPTA~\cite{EPTA:2023fyk,EPTA:2023sfo,EPTA:2023xxk}, PPTA~\cite{PPTA3-SGWB,PPTA3-pulsars,PPTA3-SMBHB} and CPTA~\cite{CPTA-SGWB},   have recently released data showing   evidence for the presence of a stochastic background of 
Gravitational Waves (GWs).  
Pulsars are  high-quality clocks generating    pulses in the radio frequencies band which arrive  at Earth with  regular and predictable arrival times. The latter are slightly perturbed by the presence of the GWs which induce correlated perturbations  between different radio pulsars,  called ``timing residuals".

The  correlations  are predicted to follow the so-called ``Hellings and Downs" (HD) correlation curve \cite{Hellings:1983fr} which describe
 the expected correlation in the timing residuals from a pair of pulsars as a function of their angular separation on the sky. It is a prediction of general relativity  for an unpolarized and isotropic GW background (see the recent Ref.  \cite{Romano:2023zhb} for a nice set of comments about the HD correlation).

 The goal of this paper is to show that the HD correlation function can be obtained by very general symmetry arguments. The key point  is the realization that  
 \begin{enumerate}

 \item  Lorentz transformations 
 between inertial observers are conformal motions of the surface of a unit two-sphere; 
 
 \item such Lorentz transformations 
can be interpreted as conformal transformations on the celestial sphere; 

\item under such transformations the HD correlation transforms in a well-defined way with a well-defined conformal weight. 
\end{enumerate}
This allows one to obtain the HD correlation  with  no explicit and cumbersome calculation.
 The symmetry arguments are also suitable  to generalize the HD correlation to the case in which three or more pulsars are correlated amd to include  additional polarization modes (vector and/or scalar) in  theories of gravity different from general relativity.

 The paper is organized as follows. In section 2 we describe the Lorentz group and show how Lorentz transformations are seen as conformal transformation in the celestial sphere.  In section 3 we obtain the generic two-point correlator on the celestial sphere and show in section 4 how the HD correlation function is recovered. Section 5 extends our computation of the overlap function to the so-called short-arm detectors. Section 6 contains our conclusions. The paper contains also three Appendices where some technical details are added.

\section{The Lorentz group, inertial observers and the celestial sphere}
Let us first review the notion of the celestial sphere of an inertial observer in Minkowski spacetime~\cite{Barut:1974dz, Penrose:1985bww,Oblak:2015qia}. 
The knowledgeable reader can skip this section.

The celestial sphere is an imaginary sphere of practically infinite radius centered at the position of an  inertial observer. 
We wish to show now that  Lorentz transformations between different inertial observers correspond to conformal transformations of the celestial sphere. First, we recall the stereographic projection and projective coordinates of the two-dimensional sphere $S^2$. The latter is defined as 
\begin{eqnarray}
S^2=\{(X,Y,Z)\in \mathbb{R}^3: X^2+Y^2+Z^2=1\}.
\end{eqnarray}
The points $(0,0,1)$ and $(0,0,-1)$ are the north $(N)$ and the 
south ($S$) poles of the sphere. The  
 $(x,y)$ plane at $z=0$ can be identified   with  
$\mathbb{C}$ by defining $\z=x+i y$. The stereographic projection 
$\z: S^2\backslash \{N\}\to \mathbb{C}$ is then the  map that 
associates to every point $(X,Y,Z)\in S^2\backslash \{N\} $
 the intersection $\z$ of the line which connects the north pole $N$ with  $(x,y,z=0$).  
 \begin{figure}
\centering
\includegraphics[scale=.5,angle=0]{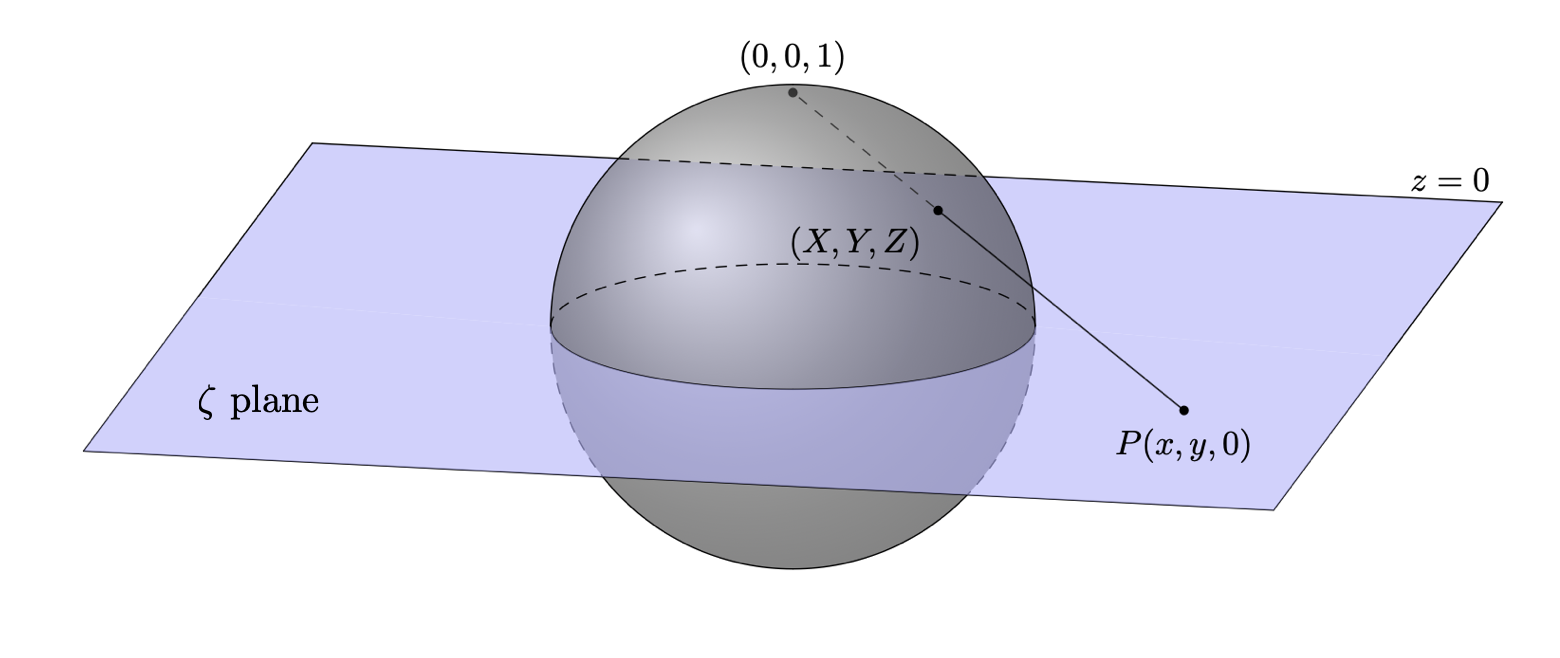}
\caption{The stereographic
projection and projective coordinates of the two-dimensional sphere $S^2$.}
\end{figure}
 In other words, we have 
 \begin{eqnarray}
 \z=x+i y=\frac{X+iY}{1-Z}.  \label{z1}
 \end{eqnarray}
 If we introduce standard spherical coordinates in $\mathbb{R}^3$ 
 \begin{eqnarray}
 X=r\sin\theta\, \cos\phi, \quad Y=r \sin\theta\,\sin\phi,\quad
 Z=r \cos\theta, 
 \end{eqnarray}
 we find that Eq. (\ref{z1}) is written as 
 \begin{eqnarray}
 \zeta(\theta,\phi)=\frac{\sin\theta}{1-\cos\theta}e^{i\phi}. 
 \end{eqnarray}
Therefore, we get 
\begin{eqnarray}
\d \zeta= -\frac{1}{1-\cos\theta}e^{i\phi}\d \theta
+i\frac{\sin\theta}{1-\cos\theta} e^{i\phi}\d \phi,
\end{eqnarray}
so that 
\begin{eqnarray}
\d \z\d \zb=\frac{1}{(1-\cos\theta)^2}\d s_2^2,  
\end{eqnarray}
where 
\be
\d s_2^2=\d \theta^2+\sin\theta^2 \d \phi^2
\ee
is the standard metric on the unit $S^2$. 
Since 
\begin{eqnarray}
\frac{1}{2}(1-\cos\theta)=\frac{1}{1+\z\zb},
\end{eqnarray}
the metric on the $S^2$ is written as 
\begin{eqnarray}
\d s_2^2=\frac{4}{(1+\z\zb)^2}\d \z\d \zb.
\label{s2}
\end{eqnarray}
Note in the above construction, the north pole is  mapped nowhere. However, by extending $\mathbb{C}$ to $\mathbb{C}\!\cup\!\{\infty\}$, the north pole is mapped to itself and therefore, we can identify the whole of $S^2$ with $\mathbb{C}\!\cup\!\{\infty\}$. Therefore, 
complex analytic automorphisms $f:S^2\to S^2$ are represented by homomorphic mappings $g: \mathbb{C}\!\cup\!\{\infty\}\to \mathbb{C}\!\cup\!\{\infty\}$. In particular, it is easy to find that 
\begin{eqnarray}
\d s_2^2(g(z))=\frac{g'\overline{g}'(1+\z\zb)^2}{(1+g\overline{g})^2}
\d s_2^2(\z),
\label{gg}
\end{eqnarray}
and therefore, complex analytic automorphisms of $S^2$ are conformal transformations. 
In fact, it can be shown that the conformal orientation-preserving automorphisms of $S^2$ form a group isomorphic to the group of all complex analytic automorphisms of $S^2$.  The latter are generated by the composition of inversion  and the linear transformation 
\begin{eqnarray}
\z\to \frac{1}{\z}, \qquad \z \to a\z+b, ~~~~(a\neq0)
\end{eqnarray}
and therefore, complex analytic automorphism of $S^2$ are the M\"obius transformations
\begin{eqnarray}
 \z\to \frac{a\z+b}{c\z +d}, 
 \end{eqnarray} 
 which form a group isomorphic to 
 $SL(2,\mathbb{C})/\mathbb{Z}_2$. 
Thus, the group of M\"obius transformations is the group of all complex analytic automorphisms of $S^2$
\begin{eqnarray}
\z\to g(\z)=\frac{a\z+b}{c\z+d}, \qquad ab-cd=1. 
\end{eqnarray}
On the other hand, it is known that the  proper orthochronous Lorentz 
transformations $O_+^\uparrow(1,3)$ is isomorphic to the M\"obius  group $SL(2,\mathbb{C})/\mathbb{Z}_2$ \cite{Barut:1974dz, Penrose:1985bww}. A convenient set of generators of the latter 
are 
\begin{eqnarray}
L_i=\frac{1}{2} \sigma_i, \qquad K_i=-\frac{i}{2}\sigma_i,
\end{eqnarray}
where $\sigma_i$ ($i=1,2,3)$ are the Pauli matrices. The generators 
$L_i,K_i$ satisfy the commutation relations of the $SL(2,\mathbb{C})$ algebra
\begin{eqnarray}
 [K_i,K_j]=-\epsilon_{ijk}L_k, \quad [L_i,L_j]=\epsilon_{ijk}L_k,\quad
 [K_i,L_j]=\epsilon_{ijk}K_k.
 \end{eqnarray} 
 Due to the aforementioned isomorphism,  one can prove that boosts in the $z$-direction in Minkowski spacetime by a hyperbolic angle $\xi$ generated by $K_3$, corresponds to the conformal (M\"obius) transformation $\z\to e^\xi\z$.
 In analogy, the generators $K_1,K_2$ of $SL(2,\mathbb{C})$ generate 
 boosts in the $x$ and $-y$ directions. Similarly, $L_1,L_2$ and $L_3$, generate rotations around the $-x$, $y$ and $-z$ directions, respectively.  
In general, there exists a map $f:SL(2,\mathbb{C})/\mathbb{Z}_2\to  O_+^\uparrow(1,3)$ given explicitly by 
\begin{eqnarray}
SL(2,\mathbb{C})\ni S=\begin{pmatrix}
a & b \\
c & d 
\end{pmatrix}
\end{eqnarray}
which implies
\begin{eqnarray}
f(S)&=&\begin{pmatrix}
\frac{1}{2}\left(|a|^2+|b|^2+|c|^2+|d|^2\right) & -{\rm Re}\left(a\overline{b}+c\overline{d}\right)&{\rm Im}\left(a\overline{b}+c\overline{}\right) &\frac{1}{2}\left(|a|^2-|b|^2+|c|^2-|d|^2\right) \\
-{\rm Re}\left(\overline{a} c+\overline{b}d\right) & {\rm Re}\left(\overline{a} d+\overline{b} c\right)& -{\rm Im}\left(a\overline{d}-b\overline{c}\right) 
&-{\rm Re}\left(\overline{a} c-\overline{b}d\right)\\
{\rm Im}\left(\overline{a} c+\overline{b}d\right)& -{\rm Im}\left(\overline{a} d+\overline{b} c\right)& {\rm Re}\left(a\overline{d}-b\overline{c}\right)&
{\rm Im}\left(\overline{a} c-\overline{b}d\right)\\
\frac{1}{2}\left(|a|^2+|b|^2-|c|^2-|d|^2\right) & -{\rm Re}\left(a\overline{b}-c\overline{d}\right)&{\rm Im}\left(a\overline{b}-c\overline{}\right) &\frac{1}{2}\left(|a|^2-|b|^2-|c|^2+|d|^2\right)
\end{pmatrix}.\nonumber\\
\label{lS}
&&
\end{eqnarray}
 One concludes that

\begin{tcolorbox}

\centerline{Lorentz transformations between  inertial observers are   conformal motions of the two-sphere.}

\end{tcolorbox}
\noindent
\subsection{Lorentz transformations and the celestial sphere}
The sphere described above is actually a spatial sphere, that is a sphere at radius $r^2=x^2+y^2+z^2$. However, we  are interested in the sphere  that an observer looks at, that is the celestial sphere. This is not a spatial sphere defined by $r=$constant, since 
  ingoing,  radial light-rays emitted by the sphere need a time $r$ to arrive  from the sphere to the observer at the origin $r= 0$, (we have set  the speed of light to unity).  
Instead, one can parametrize the  emission time of radial ingoing  light rays by the value of the advanced time coordinate $u=t+r$.
In other words, we specify events in spacetime by using  the Bondi coordinates $(u,r,\zeta)$. A sphere at constant $r >0$  and constant $u$  in Bondi coordinates, is the sphere seen by an observer at $r= 0$ and  time $u$. The celestial sphere at time $u$ is then the sphere  located at an infinite distance $r\to \infty$, and at a fixed $u$,
and it is the sphere formed by all directions towards which an observer sitting at $r=0$ may look. 

In terms of the Bondi coordinates, the metric of the flat Minkowski spacetime is written as 
\begin{eqnarray}
\d s^2=-\d u^2+2\d u \d r+r^2 \d s_2^2,
\end{eqnarray}
where $\d s_2^2$ in the metric on the unit $S^2$ given in Eq. (\ref{s2}). Isometries of Minkowski spacetime are just Lorentz transformations, which act linearly on the flat Cartesian coordinates $x^\mu$. On the other hand,  Lorentz tranformations act highly non-linear on the Bondi coordinates. However, they have a simpler form at large $r$ and in particular, one finds that under the  Lorentz transformations given in Eq. (\ref{lS}), the Bondi coordinates transform as $(u,r,\z)\to (u',r',\z')$ where 
\begin{subequations}
\begin{eqnarray}
&&r'=F(\z,\zb)\, r+\mathcal{O}(1),
\label{q1}\\
&& u'=\frac{u}{F(\z,\zb)}+\mathcal{O}(r^{-1}), \label{q2} \\
&& \z'=\frac{a\z+b}{c\z+d}+ \mathcal{O}(r^{-1}), \label{q3}
\end{eqnarray}
\end{subequations}
and  
\begin{eqnarray}
F(\z,\zb)=\frac{|a\z+b|^2+
|c\z+d|^2}{1+\z\zb}. 
\end{eqnarray}
Eq.(\ref{q3}) is clearly, to leading order, the standard Möbius transformations on $S^2$ and therefore, 

\begin{tcolorbox}

\centerline{Lorentz transformations are   conformal transformations of the celestial sphere.}
\end{tcolorbox}

\noindent
In particular, by using Eq. (\ref{lS}), we find that the 
matrix 
\begin{eqnarray}
SL(2,\mathbb{C})/\mathbb{Z}_2\ni S_0=\begin{pmatrix}
e^{-\xi/2} & 0 \\
0 & e^{\xi/2}
\end{pmatrix},
\end{eqnarray}
corresponds to the $O_+^\uparrow(1,3)$  matrix
\begin{eqnarray}
 L_{b}=\begin{pmatrix}
\cosh\xi & 0&0&-\sinh\xi \\
0 & 1&0&0\\
0&0&1&0\\
-\sinh\xi&0&0&\cosh\xi
\end{pmatrix},
\end{eqnarray}
which is nothing else than a Lorentz boost  with rapidity 
\begin{eqnarray}
e^\xi=\gamma (1-\vec{v}\cdot \wn),  \label{wnxi}
\end{eqnarray}
along the $z$-direction. In Eq. (\ref{wnxi}), the unit vector $\wn=(n_1,n_2,n_3)$  ($n_in^i=1$) is  $\widehat{e}_z$. 
Then, $F(\z,\zb)$ turns out to be
\begin{eqnarray}
 F(\z,\zb)=\gamma (1-\vec{v}\cdot \wn),\qquad \gamma=(1-v^2)^{-1/2}
 \end{eqnarray} 
and therefore, Lorentz boosts corresponds to the following transformations of the Bondi coordinates 
\begin{align}
u'&=\Omega(\wn) u,\nonumber \\
 r'&=\frac{r}{\Omega(\wn)},  \nonumber \\
\zeta'&=\Omega(\wn) \zeta,
\label{zlor}
\end{align}
where 
\begin{eqnarray}
\Omega(\wn)=\frac{1}{\gamma (1-\vec{v}\cdot \wn)},  \label{Omega}
\end{eqnarray}
is the Doppler shift factor.
It is easy to verify that Eq.(\ref{zlor}) gives rise to the aberration formula for outgoing light rays
\begin{eqnarray}
\cot\frac{\theta'}{2}=\sqrt{\frac{1-v}{1+v}}\cot\frac{\theta}{2}.
\label{abb}
\end{eqnarray}
Then by using Eq. (\ref{abb}), we find that  under a Lorentz boost in the $\wn$ direction, the metric of the celestial sphere transforms as 
\begin{eqnarray}
\d \theta^{'2}+\sin^2\theta'\d \phi^{'2}=\Omega^2(\wn) 
\Big(\d \theta^{2}+\sin^2\theta\d \phi^{2}\Big). 
\end{eqnarray}

\section{The two- and three-point Correlators on the Celestial Sphere} 
The goal of this section is two explain how a generic two-point correlator can be written on the celestial sphere.

 The projective coordinate $\z$ in $\mathbb{C}\!\cup\!\{\infty\}$ 
 can be parametrized in terms of the  unit vector $\widehat{n}$ as 
\begin{eqnarray}
\z=\frac{n^1+i n^2}{1-n^3},
\end{eqnarray}
so that 
\begin{eqnarray}
n^1=\frac{\z+\zb}{1+\z\zb},\qquad n^2=\frac{1}{i}\frac{\z-\zb}{1+\z\zb}, \qquad n^3=-\frac{1-\z\zb}{1+\z\zb},
\end{eqnarray}
and  the metric on the $S^2$ becomes 
\begin{eqnarray}
\d s_2^2=\d \wn\cdot\d\wn=\d n^i\d n^i, \qquad i=1,2,3.
\end{eqnarray}
Notice  that since the $S^2$ is formed by light rays which satisfy 
\begin{eqnarray}
{x^0} x^0-\vec{x}\cdot \vec{x}=0,
\end{eqnarray}
the unit vector can be expressed as 
\begin{eqnarray}
n^i=\frac{x^i}{x^0}, \qquad x^0=|\vec{x}|>0. 
\end{eqnarray}
Clearly,  $\wn$ transforms as a vector under space rotations, whereas, under  Lorentz boosts, which act linearly on the 
Minkowski coordinates $x^\mu$, it is easy to find, using the transformation properties of $(x^0,\vec{x})$ 
\begin{align}
    x^0{}'&=\gamma\Big(x^0-\vec{v}\cdot\vec{x}\Big), \nonumber \\
    \vec{x}\,{}'&=\vec{x}+(\gamma-1)\widehat{v}\, \vec{x}\cdot \widehat{v}-\gamma x^0\vec{v},
\end{align}
that $n^i$ transforms as 
\begin{eqnarray}
\wn \to \wn'=\Omega(\wn)\Big(\wn+(\gamma-1)\widehat{v}(\widehat{v}\cdot \wn)-\gamma \vec{v}\Big). \label{nn}
\end{eqnarray}
For an infinitesimal Lorentz boost generated by $K$,  using Eq. (\ref{Omega}) we get
\begin{eqnarray}
\delta_K \wn=-\vec{v}+(\widehat{v}\cdot\wn)\wn, \qquad 
\delta_K \Omega=\vec{v}\cdot \wn. \label{dn}
\end{eqnarray}
It is not difficult to verify   that indeed, Lorentz boosts that transform the unit vector $\wn$ as in Eq. (\ref{nn}), act like conformal transformations on  $S^2$ 
since 
\begin{eqnarray}
\d \wn\cdot \d \wn\to \d \wn'\cdot \d \wn'=\Omega^2(\wn)\d \wn\cdot \d \wn. 
\end{eqnarray}
Let us now consider the two-point correlator $C_2(\vec{x}_1,\vec{x}_2)\Big|_{S^2}=C_2(\wn_1,\wn_2)$ at equal time $t$ and for  points $\vec{x}_1$ and $\vec{x}_2$ both on the $S^2$. We will demand invariance under the Lorentz group. Rotational invariance leads to dependence on the distance 
\begin{eqnarray}
d_{12}^2=|\vec{x}_1-\vec{x}_2|^2=4 \sigma, 
\end{eqnarray}
where \begin{eqnarray}
\sigma=\frac{1}{2}
(1-\wn_1\cdot \wn_2) \label{gd1}
\end{eqnarray}
is the geodesic invariant distance between points on $S^2$. 
Therefore, we should have  

\be
C_2=C_2(\sigma,t;w),
\ee 
where $w$ is the scaling dimension of $C_2$ since Lorentz boosts are actually conformal rescalings of the $S^2$. In particular, we find that under Lorentz boosts, $\sigma $ transforms as 
\begin{eqnarray}
\sigma\to \sigma'=\Omega(\wn_1) \Omega(\wn_2) \sigma, \label{gd2}
\end{eqnarray}
or, in infinitesimal form,
\begin{eqnarray}
\delta_K \sigma=(\vec{v}\cdot\wn_1+\vec{v}\cdot \wn_2) \sigma .
\end{eqnarray}
In addition, if the scaling dimension of $C_2$ is $w$, we will have that 
\begin{eqnarray}
C_2(\sigma';w)\to C_2'(\sigma,t;w)=\Omega(\wn_1)^{-w} \Omega(\wn_2)^{-w}
C_2(\sigma;w).
\end{eqnarray}
Therefore, the change in the two-point correlator $C_2$ under Lorentz boosts is 
\begin{eqnarray}
\delta_KC_2(\sigma,t;w)=w\Big(\delta_K \Omega(\wn_1)+\delta_K \Omega(\wn_2)\Big)C_2(\sigma,t;w)+\delta_K \sigma \frac{\d C_2(\sigma,t;w)}{\d \sigma}. 
\end{eqnarray}
Hence, we find that 
\begin{eqnarray}
\delta_K C_2=\left(w C_2+\sigma \frac{\d C_2}{\d \sigma}\right) (\vec{v}\cdot\wn_1+\vec{v}\cdot \wn_2).
\end{eqnarray}
Invariance under Lorentz boosts ($\delta_KC_2=0$) then leads to 
\begin{eqnarray}
w C_2+\sigma \frac{\d C_2}{\d \sigma}=0,
\end{eqnarray}
and therefore, 
\begin{eqnarray}
C_2(\sigma;w)=\frac{C_{02}(w)}{\sigma^w}.  \label{w0}
\end{eqnarray}
In particular, for $w=0$ we get  (as a limit of Eq. (\ref{w0}))
\begin{eqnarray}
\label{clog}
C_2(\sigma,0)=\alpha_2 \ln \sigma+ \beta_{2},
\end{eqnarray}
where $\alpha_2$ and $\beta_2$ are constants.

\section{Hellings and Downs from Lorentz invariance or conformal symmetry on the celestial sphere}
We now come to the HD correlation function. The goal of this section is to demonstrate that it can be understood in terms of Lorentz transformations or conformal transformations on the celestial sphere.
Let us  consider  metric perturbations which, for the most general  GW  is a superposition of (transverse-traceless) plane waves, i.e., 
\begin{eqnarray}
\label{pw}
h_{ij}^{TT}(t,\vec{x})=\sum_{A=+,\times}\int_{-\infty}^\infty \d f\int
\d^2\wn\, h_A(f,\wn) e^A_{ij}(\wn) e^{-2\pi if(t-\wn\cdot \vec{x})},
\end{eqnarray}
where $A=+,\times$ are the polarizations of the GWs, $\wn$ their direction of propagation and $f$ their frequency. 
The polarization tensors $e^A_{ij}(\wn)$ are given by 
\begin{eqnarray}
e^+_{ij}=\widehat{ u}_i\widehat{ u}_j-\widehat{ w}_i\widehat{ w}_j,\qquad
e^\times_{ij}=\widehat{ u}_i\widehat{ w}_j+\widehat{ w}_i\widehat{ u}_j,
\end{eqnarray}
where the vectors $\wn,\widehat{u}$ and $\widehat{w}$ are explicitly given by
\begin{align}
\wn&=(-\sin\theta\cos\phi,-\sin\theta\cos\phi,-\cos\theta),\nonumber 
\\
\widehat{ u}&=(\sin\phi,-\cos\phi,0), \nonumber \\
\widehat{ w}&=(-\cos\theta\cos\phi,-\cos\theta\sin\phi,\sin\theta),
\end{align}
and depicted in Fig. 2.
\begin{figure}
\centering
\includegraphics[scale=.6]{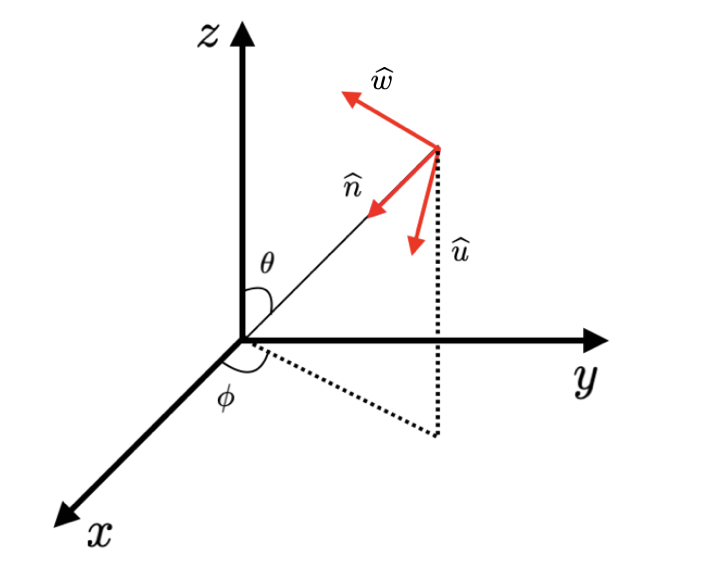}
\caption{The GW source is in the $-\wn$ direction and the GW propagates towards the origin.}
\end{figure}
In addition, we will need later  the ensemble average   
of a stationary, isotropic and unpolarized, stochastic GW background, which is characterized by 
\begin{eqnarray}
\label{hh}
\langle h^*_{A_1}(f_1,\wn_1)h_{A_2}(f_2,\wn_2)\rangle
=\delta(f_1-f_2)\frac{\delta^{(2)}(\wn_1-\wn_2)}{4\pi}\delta_{A_1A_2} \frac{1}{2}S(f_1),
\end{eqnarray}
where $S(f)$ is the spectral density of the stochastic background.


Let us now recall how the periodicity of the pulses of a pulsar is affected by the  passing GW. If the pulsar is at position 
$\vec{x}_a=d_a \wn_a$, then the change in the periodicity  (time residual) 
$$z_a=\left(\frac{\Delta T}{T}\right)_a$$  due to a GW $h_{ij}^{TT}(t,\vec{x})$  as seen by an observer at $\vec{x}=0$ is  
\begin{eqnarray}
z_a=\frac{\wn_a^i\wn_a^j }{1-\wn\cdot \wn_a}
\bigg(h_{ij}^{TT}(t,\wn,\vec{x}=0)-
h_{ij}^{TT}(t-d_a/c,\wn_a,\vec{x}_a)\bigg).
\end{eqnarray}
The term in the parenthesis represents the difference between the metric perturbation which arrive at the position of the observer at $\vec{x}=0$ and the position of the  pulsar at $\vec{x}_a$. As it is standard in any similar consideration,  we will ignore the pulsar metric perturbation \cite{Maggiore:2018sht} and  write instead
\begin{eqnarray}
z_a(t,\wn)=\frac{\wn_a^i\wn_a^j }{1-\wn\cdot \wn_a}
h_{ij}^{TT}(t,\wn,\vec{x}=0).
\end{eqnarray}
A stochastic GW background can be constructed by  the  superposition of plane GW of all possible  frequencies  from all possible directions. Therefore, the change in the periodicity of the pulsar due to a stochastic GW background  can be expressed  as 
\begin{eqnarray}
\label{dd}
z_a(t)=\wn_a^i\wn_a^j \sum_{A} \int_{-\infty}^{+\infty} \d f \int \frac{\d^2 {\wn}}{1-\wn\cdot \wn_a} h_{A}(f,\wn) e_{ij}^A(\wn)\,e^{-2\pi i f t},
\end{eqnarray}
where Eq. (\ref{pw}) has been used. Then by using Eq. (\ref{hh}), we find that the correlator of time residuals  $\langle z_a z_b\rangle$ of two pulsars at directions $\wn_a$ and $\wn_b$ is given by
\begin{eqnarray}
 \langle z_a z_b\rangle=\frac{1}{2}\int_{-\infty}^\infty \d f \, S(f) \, \Gamma_{ab}(\wn_a,\wn_b)
  \end{eqnarray} 
  where we have defined the overlap function
  \begin{eqnarray}
  \Gamma_{ab}(\wn_a,\wn_b)=
 \int\frac{\d^2 \wn}{4\pi}  \frac{\wn_a^i\wn_a^j}{1-\wn\cdot\wn_a}
 \frac{\wn_b^k\wn_b^l}{1-\wn\cdot\wn_b}\sum_{A=+,\times}e^A_{ij}(\wn)
 e^A_{kl}(\wn). \label{cC}
   \end{eqnarray} 
We are interested in the transformation properties of the correlator 
$\Gamma_{ab}$ in Eq.(\ref{cC}) under Lorentz transformations. 
We have seen that three-dimensional  rotational invariance requires that the correlator is a function of $\sigma_{ab}$ only  

\begin{equation}
    \Gamma_{ab}=\Gamma_{ab}(\sigma_{ab}),
\end{equation}
that is a function of the angular separation of the pulsars.
It remains to determine the transformation properties of $\Gamma_{ab}$ under Lorentz boosts.  We will consider  a Lorentz boost of velocity 
$\vec{v}$ in the direction of the GW, i.e., $\vec{v}=\wn$. Notice that this particular choice does not spoil the generality of the following analysis, since we need to integrate over all directions 
$\wn$ to account for a stochastic GW background. This is equivalent to keeping $\wn$ fixed and boosting in all possible directions.
 To determine the transformation law of $\Gamma_{ab}$, we recall 
  Eqs.(\ref{nn}), (\ref{gd1}) and (\ref{gd2}) so that
 \begin{align}
 \wn_a^{\prime i}=\Omega(\wn_a)\Big(\wn_a^i+\widehat{v}^i(\widehat{v}\cdot\wn_a)(\gamma-1)-\gamma v^i\Big), \label{nnn}
 \end{align}
 and 
 \begin{align}
 1-\wn'\cdot \wn_a'=\Omega(\wn)\Omega(\wn_a)(1-\wn\cdot \wn_a).  \label{n;n;}
 \end{align}
 The polarization tensors $e^A_{ij}$  are invariant under Lorentz boosts along the direction of propagation. Indeed, in this case, 
 $(\widehat{u},\widehat{w})$ are invariant and so are the polarization tensors. In the Newman-Penrose formalism, $(\widehat{u},\widehat{w})$ span the $(m^\alpha,\overline{m}^a$) plane, which is invariant under Lorentz boosts in the $(\ell^a,n^a)$ directions. 
Thus, using Eq.(\ref{nnn}) we find that 
\begin{eqnarray}
\wn_a^{'i}\wn_a^{'j}e^{A'}_{ij}(\wn')=\wn_a^{i}\wn_a^{j}e^{A}_{ij}(\wn).
\end{eqnarray}
Under a Lorentz boost, the volume element on the $S^2$  in 
Eq. (\ref{cC}) transforms as 
\begin{eqnarray}
\d^2 \wn\to \d^2 \wn'\Omega^2(\wn). 
\end{eqnarray} 
The amplitude of the GW is also invariant as follows from the transformation of the Weyl scalar $\Psi_4$ under Lorentz boosts~\cite{Janis:1965tx}
\begin{eqnarray}
\Psi_4'=\Omega^{-2}(\wn) \Psi_4. 
\end{eqnarray}
Since~\cite{Teukolsky:1973ha}
\begin{eqnarray}
\Psi_4=\frac{\omega^2}{2}\Big(h_+^2-i h_\times^2\Big),
\end{eqnarray}
and $\omega=2\pi f$ transforms as 
\begin{eqnarray}
\omega'=\Omega^{-1}(\wn) \omega, \label{ff}
\end{eqnarray}
we get that the amplitudes $h_A$ are invariant under Lorentz boosts. 
Therefore, for a stationary, isotropic and unpolarized stochastic GW background,   the ensemble average in Eq. (\ref{hh}) is invariant
\begin{eqnarray}
\langle h^*_{A_1}(f_1,\wn_1)h_{A_2}(f_2,\wn_2)\rangle
=\langle h^{*'}_{A_1}(f_a',\wn_1')h'_{A_2}(f_2',\wn_2')\rangle
\end{eqnarray}
or, in other words, 
\begin{eqnarray}
\delta(f_1'-f_2')\frac{\delta^{(2)}(\wn_1'-\wn_2')}{4\pi}\delta_{A_1A_2} \frac{1}{2}S'(f_1')= \delta(f_1-f_2)\frac{\delta^{(2)}(\wn_1-\wn_2)}{4\pi}\delta_{A_1A_2} \frac{1}{2}S(f_1)
\end{eqnarray}
Since $\delta^{(2)}(\wn_1-\wn_2)$ and $\delta(f_1-f_2)$ transform as 
\begin{eqnarray}
\delta^{(2)}(\wn_1-\wn_2)\to \Omega(-\wn_1)^2\delta^{(2)}(\wn_1-\wn_2), \qquad \delta(f_1-f_2)\to \Omega(\wn_1) \delta(f_1-f_2),
\end{eqnarray}
respectively, we get that the spectral density of the stochastic background transforms as 
\begin{eqnarray}
S'(f_1)\to \Omega(\wn_1) S(f_1). 
\end{eqnarray}
and therefore, since $\d f$ transforms as 
\begin{equation}
    \d f\to \d f'=\Omega(\wn)\d f, \label{df}
\end{equation}
we find that 
$\d f S(f)$ is invariant
\begin{eqnarray}
\d f' S'(f')=\Omega(\wn)^{-1} \d f \, \Omega(\wn) S(f)=\d f S(f). 
\end{eqnarray}
This can also be deduced from the relation 
\begin{eqnarray}
\langle h_{ij}h^{ij}\rangle=4\int_0^\infty \d f \, S(f).
\label{fS}
\end{eqnarray}
Since the left-hand side of Eq. (\ref{fS}) is invariant, so is the right-hand side. 

Putting all together, it is easy to verify that overlap function transforms under Lorentz boosts  as  
\begin{align}
\Gamma_{ab}\to \Gamma'_{ab}&= 
 \int\frac{\d^2 \wn}{4\pi}  \frac{\wn_a^i\wn_a^j}{1-\wn\cdot\wn_a}
 \frac{\wn_b^k\wn_b^l}{1-\wn\cdot\wn_b} \Omega(\wn_a)\Omega(\wn_b)\sum_{A=+,\times}e^A_{ij}(\wn)
 e^A_{kl}(\wn) .
\label{ccc}
\end{align} 
 However, this transformation has fixed points since there are Lorentz boosts that leave  $\Gamma_{ab}$ invariant. Indeed,  since 
\begin{eqnarray}
\Omega(\widehat{v})\Omega(-\widehat{v})=1,
\end{eqnarray}
for pulsars located at opposite directions, $\wn_a$ and $\wn_b=-\wn_a$, a Lorentz boost along their direction (so that $\widehat{v}=\pm \wn_a$), leaves  their correlator $\Gamma_v=\Gamma_{ab}(\pm v,\mp v)$ invariant
\begin{eqnarray}
\Gamma_{v}'=\Gamma_{v}.
\end{eqnarray}  
Then, the reduced overlap function  $\Gamma_{ab}^-=\Gamma_{ab}-\Gamma_v$ transforms as in Eq. (\ref{ccc}) without fixed points and therefore, 
\begin{eqnarray}
\overline{\Gamma}_{ab}=\frac{\Gamma^-_{ab}}{\sigma_{ab}}=
\int\frac{\d^2 \wn}{4\pi}  \frac{\wn_a^i\wn_a^j}{1-\wn\cdot\wn_a}
 \frac{\wn_b^k\wn_b^l}{1-\wn\cdot\wn_b} \frac{1}{\sigma_{ab}}\sum_{A=+,\times}e^A_{ij}(\wn)
 e^A_{kl}(\wn) 
\end{eqnarray}
is invariant under Lorentz boosts 
\begin{eqnarray}
\overline{\Gamma}'_{ab}=\overline{\Gamma}_{ab}.
\end{eqnarray}
Thus, allowing logarithmic scaling violation (see  Appendix A for a technical explanation), by using  Eq. (\ref{clog}), we find
\begin{eqnarray}
\overline{\Gamma}_{ab}=\alpha \, \ln \sigma_{ab}+\beta. 
\end{eqnarray}
from where we get  
 the functional form of the HD curve

 \vspace{0.2cm}
 \begin{center}
     \label{HD1}
\tcbox[nobeforeafter]{$\Gamma_{ab}= \alpha\,\sigma_{ab}\ln\sigma_{ab}+\beta \,\sigma_{ab}+\Gamma_v$.}
\end{center}
\vspace{-1cm}
\be\ee
\vskip 0.5cm
\subsection{Recovering the Hellings and Downs correlation}
Our goal now  is to determine the constants $\alpha$, $\beta$ and $\Gamma_v$ of Eq. (\ref{HD1}) and to recover the HD correlation function.
For this we need two conditions:

\begin{enumerate}
    \item  PTA detectors operate in the ``long-arm" (or short-wavelength) limit in which the  distance between the pulsars and the earth is much  longer than the wavelengths of the GWs; this induces 
    a well-defined    
relation of the correlators for pulsars in the same and in opposite directions; 

\item  the vanishing of the integral of the  two-point correlator over the whole celestial sphere basically due to the isotropy of the stochastic GW background and its  traceless and transverse properties.
\end{enumerate}
First, as shown in Appendix B, by using the fact that the two-point correlator of two pulsars separated by $\pi$ ($\sigma_{ab}=1$) is half that for $0$  angular separation
($\sigma=0$), we get that 
\begin{eqnarray}
 \label{HD11}
\Gamma_{ab}= \alpha\,\sigma_{ab}\ln\sigma_{ab}+\beta \,\sigma_{ab}-2\beta.
\end{eqnarray}
Secondly,  as shown in Appendix C, one can prove that 
the integral of the two-point function over the whole celestial sphere should vanish and therefore 
\begin{eqnarray}
 \label{HD2}
\int \d^2 \Omega_2 \,\Gamma_{ab}=-4\pi\int_{0}^1 \d \sigma_{ab} \Big(\alpha\,\sigma_{ab}\ln\sigma_{ab}+\beta \,\sigma_{ab}-2\beta\Big)=\pi(\alpha+6 \beta)=0,
\end{eqnarray} 
This leads to
\begin{eqnarray}
\beta=-\frac{\alpha}{6}.
\end{eqnarray}
Knowing that the overall normalization, that is the coefficient $\alpha$ is arbitrary \cite{Romano:2023zhb}, setting $\alpha=1$,  standard form of the  HD curve is recovered

 \vspace{0.2cm}
 \begin{center}
     \label{HD3}
\tcbox[nobeforeafter]{$\Gamma_{ab}= \sigma_{ab}\ln\sigma_{ab}-\dfrac{1}{6} \,\sigma_{ab}+\dfrac{1}{3}$,}
\end{center}
\vspace{-1cm}
\be\ee
 is obtained. Choosing $\alpha=3/2$, one would have got the alternative, but equivalent, HD function used in the literature

 \vspace{0.2cm}
 \begin{center}
     \label{HD31}
\tcbox[nobeforeafter]{$\Gamma_{ab}= \dfrac{3}{2}\sigma_{ab}\ln\sigma_{ab}-\dfrac{1}{4} \,\sigma_{ab}+\dfrac{1}{2}$.}
\end{center}
\vspace{-1cm}
\be\ee

 \subsection{Going beyond the  Hellings and Downs correlation: the three-point correlator}
Conformal symmetry on the celestial sphere can also specify (up to a multiplicative constant) the three-point correlator. Indeed, applying similar considerations as in the previous section,   the three-point correlator $C_3(\vec{x}_1,\vec{x}_2,\vec{x}_3)\Big|_{S^2}=C_3(\wn_1,\wn_2,\wn_3)$ is again,  due to rotational symmetry, a function of the geodesic invariant distances $\sigma_{ij}=|\vec{x}_i-\vec{x}_j|^2/4$, ($i,j=1,2,3$) so that $C_3=C_3(\sigma_{12},\sigma_{23},\sigma_{13})$. 
Let us further assume that under   Lorentz boosts (conformal transformations), the three-point correlator $C_3$ of fields of dimension $w$ transforms as 
\begin{align}
\delta_K C_3&=w\Big(\delta_K \Omega(\wn_1)+\delta_K \Omega(\wn_2)+\delta_K \Omega(\wn_3)\Big)C_3+\delta\sigma_{12}\frac{\partial C_3}{\partial\sigma_{12}}+\delta\sigma_{23}\frac{\partial C_3}{\partial\sigma_{23}}+\delta\sigma_{13}\frac{\partial C_3}{\partial \sigma_{13}}\nonumber \\
&=w \vec{v} \cdot (\wn_1+\wn_2+\wn_3)C_3+ \vec{v} \cdot (\wn_1+\wn_2)\sigma_{12}\frac{\partial C_3}{\partial\sigma_{12}}+
\vec{v} \cdot (\wn_2+\wn_3)\sigma_{23}\frac{\partial C_3}{\partial\sigma_{23}}\nonumber \\
&+
\vec{v} \cdot (\wn_1+\wn_3)\sigma_{13}\frac{\partial C_3}{\partial\sigma_{13}}.
\end{align}
Then, since  invariance under Lorentz boosts  $(\delta_K C_3=0)$ should hold for any $\wn_i$ and any $\vec{v}$, one is led to the equations
\be
wC_3=2\sigma_{12}\frac{\partial C_3}{\partial\sigma_{12}}=2\sigma_{23}\frac{\partial C_3}{\partial\sigma_{23}}=2\sigma_{13}\frac{\partial C_3}{\partial\sigma_{13}},
\ee
which are solved by 
\be
C_3(\sigma_{12},\sigma_{23},\sigma_{13},w)=\frac{C_{03}(w)}{\Big(\sigma_{12}\sigma_{23}\sigma_{13}\Big)^{\frac{w}{2}}}.
\ee
Again, for $w=0$, the three-point correlators $C_3(\sigma_{12},\sigma_{23},\sigma_{13},w)=C_3(\sigma,w)$, turns out to be
\be
C_3(\sigma,0)=\alpha_3\,  \ln \Big(\sigma_{12}\sigma_{23}\sigma_{13}\Big)+\beta_3,
\label{abc1}
\ee
where $\alpha_3$ and $\beta_3$ are constants.
To proceed, let us assume that there exists some sort of non-Gaussianity in the stochastic gravitational wave background. 
 In this case, 
 the three-point correlator of time residuals $C_{abc}=\langle z_a z_bz_c\rangle$ will be given by
 \begin{align}
 C_{abc}=&\int\d f_1\d f_2\d f_3 \int 
 \frac{\d^2 \wn_1}{4\pi}\frac{\d^2 \wn_2}{4\pi}\frac{\d^2 \wn_3}{4\pi}\frac{\wn_a^i\wn_a^j\wn_b^k\wn_b^l\wn_c^m\wn_c^n}{(1-\wn_1\cdot \wn_a)(1-\wn_2\cdot \wn_b)(1-\wn_3\cdot\wn_c)} e^{-2\pi i(f_1+f_2+f_3)t}\nonumber \\
 &\hspace*{1cm}\langle h^A(f_1,\wn_1)h^{A'}(f_2,\wn_2)h^{A''}(f_3,\wn_3)\rangle \sum_{A,A',A''=+,\times}\epsilon^A_{ij}(\wn_1)\epsilon^{A'}_{kl}(\wn_2)\epsilon^{A''}_{mn}(\wn_3).
 \label{3pt}
 \end{align}
 We are interested in the transformation properties of the above correlator under Lorentz boosts, that is conformal transformations of the celestial sphere. By rotational symmetry, the correlator in Eq. (\ref{3pt}) will depend on the invariants $\sigma_{ab},~\sigma_{bc}$ and $\sigma_{ac}$. 
 By using Eqs. (\ref{n;n;}), (\ref{ff}) and (\ref{df}), it  can easily be verified that the three-point correlator transforms under Lorentz boosts as
\begin{align}
\delta_KC_{abc}&=-\vec{v} \cdot (\wn_a+\wn_b+\wn_c)C_{abc}+ \vec{v} \cdot (\wn_a+\wn_b)\sigma_{ab}\frac{\partial C_{abc}}{\partial\sigma_{ab}}+
\vec{v} \cdot (\wn_b+\wn_c)\sigma_{bc}\frac{\partial C_{abc}}{\partial\sigma_{bc}}\nonumber \\
&+
\vec{v} \cdot (\wn_a+\wn_c)\sigma_{ac}\frac{\partial C_{abc}}{\partial\sigma_{ac}} + \Omega_{abc},
\end{align} 
where 
\begin{eqnarray}
\Omega_{abc}=&-\int\d f_1\d f_2\d f_3 \int 
 \frac{\d^2 \wn_1}{4\pi}\frac{\d^2 \wn_2}{4\pi}\frac{\d^2 \wn_3}{4\pi}\dfrac{2(v^i\wn_a^j\wn_b^k\wn_b^l\wn_c^m\wn_c^n+\mbox{cyclic})}{(1-\wn_1\cdot \wn_a)(1-\wn_2\cdot \wn_b)(1-\wn_3\cdot\wn_c)} e^{-2\pi i(f_1+f_2+f_3)t}\nonumber \\
 &\hspace*{1cm}\langle h^A(f_1,\wn_1)h^{A'}(f_2,\wn_2)h^{A''}(f_3,\wn_3)\rangle \sum_{A,A',A''=+,\times}\epsilon^A_{ij}(\wn_1)\epsilon^{A'}_{kl}(\wn_2)\epsilon^{A''}_{mn}(\wn_3)\nonumber \\
 &=-\int\d f_1\d f_2\d f_3 \int 
 \frac{\d^2 \wn_1}{4\pi}\frac{\d^2 \wn_2}{4\pi}\frac{\d^2 \wn_3}{4\pi}\dfrac{6v^i\wn_a^j\wn_b^k\wn_b^l\wn_c^m\wn_c^n}{(1-\wn_1\cdot \wn_a)(1-\wn_2\cdot \wn_b)(1-\wn_3\cdot\wn_c)} e^{-2\pi i(f_1+f_2+f_3)t}\nonumber \\
 &\hspace*{1cm}\langle h^A(f_1,\wn_1)h^{A'}(f_2,\wn_2)h^{A''}(f_3,\wn_3)\rangle \sum_{A,A',A''=+,\times}\epsilon^A_{ij}(\wn_1)\epsilon^{A'}_{kl}(\wn_2)\epsilon^{A''}_{mn}(\wn_3).
 \label{omega}
\end{eqnarray}
 In the last equality in Eq. (\ref{omega}),  the invariance of $C_{abc}$ under $n_i\to n_j$ ($i,j=1,2,3$)  has been used. 
As in the case of the two-point correlator, by defining 
 \begin{eqnarray}
\overline{C}_{abc}=\frac{C_{abc}}{(\sigma_{ab}\sigma_{bc}\sigma_{ac})^{1/2}}, 
 \end{eqnarray}
 and taking $\widehat{v}=\wn_1$, we find  that $\Omega_{abc}=0$. 
 In addition, it can be easily verified that $\overline{C}_{abc}$ transforms like a three-point correlator for fields of dimension $w=0$, and therefore, according to Eq. (\ref{abc1}), it will be given by 
 \begin{eqnarray}
\overline{C}_{abc}=\alpha_3\ln(\sigma_{ab}\sigma_{bc}\sigma_{ac})+\beta_3.
 \end{eqnarray}
 This specifies the three-point correlator  of the time residuals to be 
 \begin{eqnarray}
 C_{abc}=(\sigma_{ab}\sigma_{bc}\sigma_{ac})^{1/2}\, 
\bigg(\alpha_3\ln(\sigma_{ab}\sigma_{bc}\sigma_{ac})+ \beta_3\bigg),
\label{3ptp}
 \end{eqnarray}
 It should be noted that both the two- and three-point correlators have the form of the corresponding correlators of a logarithmic conformal field theory as discussed in Appendix A.

The Eq. (\ref{3ptp})  is our generic result, where the coefficients depend on the non-Gaussian nature of the tensor modes. However, even if the  GW source signal is intrinsically non-Gaussian and with a non-vanishing three-point correlator,  one has to account for the fact that the stochastic signal measured by the PTA collaborations  is a sum of the superposition of GWs  from a large number of independent sources. This suppresses  the non-Gaussianity of the time residuals, since it destroys the phase coherence needed to have a nonvanishing non-Gaussianity. Indeed, consider that the GW propagates in a perturbed universe and travel for cosmological distances. Each GW will pick up a Shapiro time delay from the time of emission $t_e$

\begin{equation}
  \delta t=2\int_{t_e}^t{\rm d}t'\Psi(\vec x+\wn(t'-t),t'),  
\end{equation}
where $\Psi$ is the gravitational potential along the trajectory and $t_e$ the emission time. When interfering with the light from the pulsars, the GW has therefore acquired a
phase shift of  the order of $f\delta t$ compared to the propagation in a homogeneous universe. This would  not a problem were    the
phase shift  the same for all the GW measurements. However, if they
vary, the average correlation of waves at three points
would pick up a factor of 

\begin{equation}
    \Big< {\rm exp}(i\sum_i f_i\delta t_i)\Big>\sim {\rm exp}\left(-\sum_i f^2_i\langle \delta t^2\rangle\right))\sim 
{\rm exp}\left(-10^{-8}\sum_i f^2_i t^2\right),
\end{equation}
which gives an exponential suppression of the three-point correlator of the time residuals when evaluated at the age of the universe  \cite{Bartolo:2018evs,Bartolo:2018rku}, unless one considers flattened triangle configurations in frequency space \cite{Powell:2019kid}. Eventually, one might also hope to escape the suppression if the GWs are produced within the same structure, e.g. by clustered binaries of supermassive black holes.

\section{Recovering the overlap function for an array of “short-arm” LIGO-like detectors}
The Hellings and Downs correlation is obtained in the so-called long-arm approximation, when the wavelengths of the GWs are much smaller than the distance between the earth and the pulsars. In the opposite limit, the so-called short-arm, in which the wavelengths are much larger than the arms of the detectors, e.g., LIGO and Virgo, the overlap function
is also easily recovered by our symmetry arguments as follows. 

The time residual is written in the short-arm approximation as   
\begin{eqnarray}
\label{dd1}
z_a(t)= \int_{-\infty}^{+\infty} \d f\, f \int \d^2 {\wn}\,\,
\wn_a^i\wn_a^j \sum_{A} h_{A}(f,\wn) \, e_{ij}^A(\wn)\,e^{-2\pi i f t} , 
\end{eqnarray}
and therefore, 
the correlator $\langle z_a z_b\rangle$ turns out to be in this case
\begin{eqnarray}
 \langle z_a z_b\rangle=\frac{1}{2}\int_{-\infty}^\infty \d f \, S(f)  \, \Gamma_{ab}^f(\wn_a,\wn_b),
  \end{eqnarray} 
  where the 
   overlap function is given by
  \begin{eqnarray}
  \Gamma_{ab}^f(\wn_a,\wn_b)=
 \int\frac{\d^2 \wn}{4\pi}  \wn_a^i\wn_a^j
 \wn_b^k\wn_b^l\, f^2\sum_{A=+,\times}e^A_{ij}(\wn)
 e^A_{kl}(\wn). \label{cC1}
   \end{eqnarray} 
   Repeating the steps of section 4, we define 
 ${\Gamma^{f-}_{ab}}=\Gamma_{ab}^f-\Gamma_v^f$, which transforms under Lorentz boost as 
 \be
\delta_K {\Gamma^{f-}_{ab}}=2\bigg((-v^i+\vec{v}\cdot \wn_a \wn_a^i)\wn_a^j\wn_b^l\wn_b^k +(-v^i+\vec{v}\cdot \wn_b\wn_b^i)\wn_b^j\wn_a^k\wn_a^l\bigg)
\int\frac{\d^2 \wn}{4\pi} f^2\sum_{A=+,\times}e^A_{ij}(\wn)
 e^A_{kl}(\wn),
 \ee
 where the transformation of Eq.(\ref{dn}) has been used. Then, by using the fact that, since $\Gamma_{ab}^f$ is quadratic in $\wn_a$, it is invariant under $\wn_a\to -\wn_a$
 or, equivalently, under  
$\sigma_{ab}\to 1-\sigma_{ab}$, we get 
\be
\delta_K {\Gamma^{f-}_{ab}}= \vec{v}\cdot (\wn_a +\wn_b) {\Gamma^{f-}_{ab}}-\frac{1}{2}\left(
\delta \wn_a\cdot \nabla_{{\wn_a}} {\Gamma^{f-}_{ab}}
+\delta \wn_b\cdot \nabla_{{\wn_b}} {\Gamma^{f-}_{ab}}\right).
\ee
Therefore, invariance under Lorentz boosts ($\delta_K {\Gamma^f_{ab}}^-=0$) leads to 
\be
\vec{v}\cdot (\wn_a +\wn_b)\bigg(2 {\Gamma^{f-}_{ab}}-
\sigma_{ab}\frac{\d {\Gamma^{f-}_{ab}} }{\d \sigma_{ab}}\bigg)=0,
\ee
which is solved by 
\begin{eqnarray}
{\Gamma^{f-}_{ab}}=\alpha\,\sigma_{ab}^2~~~~{\rm and}~~~~
\Gamma_{ab}^f=\alpha \, \sigma^2_{ab}+\Gamma_v^f.
\end{eqnarray}
Implementing the $\sigma_{ab}\to 1-\sigma_{ab}$ symmetry of  the  overlap function we  find 
\begin{eqnarray}
{\Gamma^f_{ab}}=\frac{\alpha}{2}\Big(\sigma_{ab}^2+(1-\sigma_{ab})^2\Big)+\Gamma_v^f. \label{ffl}
\end{eqnarray}
In addition, the identity
\begin{eqnarray}
\int \d^2 \wn_a \, \wn_a^i\wn_a^i=\frac{4\pi}{3} \delta^{ij},
\end{eqnarray}
leads to the condition 
\begin{eqnarray}
\int \d^2 \wn_a\, \Gamma_{ab}^f=0, \label{ppo}
\end{eqnarray}
as the polarization tensors are traceless. Using Eq. (\ref{ffl}) in  Eq. (\ref{ppo}) we find that $\Gamma_v^f=-\alpha/3$. Therefore, the final expression for the overlap function in the short-arm approximation turns out to be

\vspace{0.2cm}
 \begin{center}
     \label{HD30}
\tcbox[nobeforeafter]{$\Gamma^f_{ab}=\dfrac{\alpha}{2}\left(\sigma_{ab}^2+(1-\sigma_{ab})^2-\dfrac{2}{3}\right)=\dfrac{\alpha}{6}P_2(\wn_a\cdot\wn_b)$,}
\end{center}
\vspace{-1cm}
\be\ee
with $P_2$ the $\ell=2$ Legendre polynomial $P_\ell$. It provides the same value at opposite angles. The coefficient $\alpha$ is an arbitrary parameter, which can be set to 3 to reproduce the value reported in the literature \cite{Romano:2023zhb}.

 \section{Conclusions}
We have shown that the HD correlation function can be recovered by simple symmetry arguments once it is realized that Lorentz symmetries act as conformal symmetries on the celestial sphere and that the HD correlation function has well-defined conformal transformation properties.

We believe that this result is useful to describe variations of the standard HD function, for instance when there are 
 additional polarization modes (vector and/or scalar) that may arise in alternative-gravity theories \cite{Liang:2021bct,Liang:2023ary} for which a different angular correlation functions are expected. Anisotropy, linear, or circular polarization in the stochastic GW background \cite{Gair:2015hra} gives rise to extra  structure in the two-point correlation function and cannot  be written simply in terms of the angular separation of the two pulsars. One relevant question  we would like to investigate is if a deformation of the conformal field theory on the celestial sphere can help in describing these cases too.

\vskip 0.3cm
\noindent
\centerline{{\bf Acknowledgments}} 
\vskip 0.1cm
\noindent
We thank V. De Luca, A. Mitridate, K. Papadodimas, J. Romano, D. Racco and G. Tasinato for useful discussions. A.R. is supported by the Boninchi Foundation for the project ``PBHs in the Era of GW Astronomy''.
\vskip 1cm
\appendix
\section{The logarithmic origin of the Hellings and Downs function}
  Logarithmic terms  appear in the two-point function for conformal field theories in two dimensions. However, we should also notice, that  such terms naturally appear  in logarithimc conformal theories \cite{Rozansky:1992rx,Gurarie:1993xq}. The latter have been employed in the description of percolation \cite{Saleur:1998hq},
 the quantum Hall effect \cite{Gurarie:1997dw}, planar  magnetohydrodynamics \cite{Flohr:1996vc}, AdS/CFT  correspondence \cite{Ghezelbash:1998rj,Kogan:1999bn} whereas they have also  been discussed in a cosmological context in Ref. \cite{Kehagias:2012pd}. 
These theories are conformal invariant in a broader sense. Usually, a generic field $\phi(\vec x)$ of dimension $w$ transforms under  dilations
as 
\begin{eqnarray}
[D,\phi(\vec x)]=\left(x^i\partial_i+w\right)\phi(\vec x).
\end{eqnarray}
 However, in the presence of second field $\Phi$ of the same
 dimension $w$, we may have under dilations the transformation 
 \begin{eqnarray}
 i[D,\Phi_\alpha(x)]=\left(x^i\partial_i\delta_\alpha^\beta+\Delta_\alpha^\beta\right)\Phi_\alpha(x), ~~~~a=1,2.
 \end{eqnarray}
where $\Phi_\alpha=(\phi,\Phi)$  is the logarithmic pair, and $\Delta_\alpha^\beta$ is a $2\times2$ matrix.
In the standard case, where $\Delta_a^b=w\,\delta_a^b$, the two-point correlator is 
\begin{eqnarray}
\langle\Phi_\alpha(0)\Phi_\beta(\vec{x})\rangle=c\,\delta_{\alpha\beta}\, 
 |\vec{x}|^{-2w}.
\end{eqnarray}
However, in the most general case, the matrix $\Delta_\alpha^\beta$ need not be diagonal, and it can be brought to its  Jordan canonical form  
\be
{\bf \Delta}=\left(\begin{array}{cc}
                   w&0\\
                   1&w
                  \end{array} \right).
\ee 
Then, the two-point correlators  
$G_{\alpha\beta}(\vec{x},\vec{y})=\langle \Phi_\alpha(\vec{x})\Phi_\beta(\vec{y})\rangle$ are  found by solving the corresponding Ward identities, which are explicitly written as \cite{Kehagias:2012pd}
\begin{align}
&\left(2 w+ x\frac{\partial}{\partial x}\right)G_{12}+G_{11}=0,  \qquad \left(2 w+ x\frac{\partial}{\partial x}\right)G_{11}=0
,\nonumber \\
&\left(2 w+ x\frac{\partial}{\partial x}\right)G_{22}+2G_{12}=0,
\qquad \Delta_{\alpha\beta}G_{\beta\gamma}=G_{\alpha\beta}\Delta_{\gamma\beta},
\end{align}
where $x=|\vec{x}|$. The solution of the above equations turns out to be 
\begin{eqnarray}
&&G_{12}= -\frac{c}{2}{x^{2w}}, ~~~~~~~~G_{11}=0, \nonumber \\
&&G_{22}= 
\frac{c}{x^{2w}}
\left( \ln x+\frac{\beta}{c}\right).
\label{ggg}
\end{eqnarray}
Therefore, for $w=-1/2$ and  $G_{22}=\Gamma^-_{ab}$, we find that 
\begin{eqnarray}
\Gamma_{ab}= c\,  \sigma_{ab}\ln\sigma_{ab}+\beta \,\sigma_{ab}+\Gamma_v,
\end{eqnarray}
i.e., the functional form of the Hellings and Downs curve.

 A similar treatment of the three-point function for fields at $\vec{x}_1,\vec{x}_2$ and $\vec{x}_3$ leads to the non-zero correlators \cite{Kehagias:2012pd}
\begin{eqnarray}
&&G_{122}=\langle \phi(\vec{x}_1)\Phi(\vec{x}_2)\Phi(\vec{x}_3)\rangle = \frac{c_1}{2}\, x_{12}^{-w}x_{23}^{-w}x_{13}^{-w}, \\
&& G_{222}= \langle \Phi(\vec{x}_1)\Phi(\vec{x}_2)\Phi(\vec{x}_3)\rangle = c_1 x_{12}^{-w}x_{23}^{-w}x_{13}^{-w}\Big( \ln (x_{12}x_{23}x_{13})+a\Big),
\end{eqnarray}
where $x_{ij}=|\vec{x}_i-\vec{x}_j|$.  Again for conformal dimension $w=-1/2$, we get the three--point correlator of Eq. (\ref{3ptp}).

\section{The correlator for collinear pulsars}

By using the sum helicity rule \cite{Anninos:2019nib}
\begin{align}
\sum_{A=+,\times}\epsilon^A_{ij}(\wn)\epsilon^A_{kl}(\wn)=&
\delta_{ik}\delta_{jl}+\delta_{il}\delta_{jk}-\delta_{ij}\delta_{kl}-\delta_{ik}\wn_j\wn_l-\delta_{il}\wn_j\wn_k\nonumber\\
&-\delta_{jk}\wn_i\wn_l-\delta_{jl}\wn_i\wn_k
+\delta_{ij}\wn_k\wn_l+\delta_{kl}\wn_i\wn_j+\wn_i\wn_j\wn_k\wn_l,
 \end{align} 
 we find that 
 \begin{eqnarray}
\label{wwww}\wn_a^i\wn_a^j\wn_a^k\wn_a^l\bigg(\sum_{A=+,\times}\epsilon^A_{ij}(\wn)\epsilon^A_{kl}(\wn)\bigg)=(1-\wn\cdot \wn_a)^2(1+\wn\cdot\wn_a)^2.
 \end{eqnarray}
 Then, it turns out that the two-point correlators 
 $C_2(0)=\Gamma_{ab}(\sigma)\Big|_{\sigma=0}$  
 for 
 $\wn_b=\wn_a $ and $C_2(1)=\Gamma_{ab}(\sigma)\Big|_{\sigma=1}$  
 for $\wn_b=-\wn_a$
 are given by
\begin{align}
 C_{2}(0)=
 \int\frac{\d^2 \wn}{4\pi}  \frac{\wn_a^i\wn_a^j\wn_b^k\wn_b^l}{(1-\wn\cdot\wn_a)^2}\left(
 \sum_{A=+,\times}e^A_{ij}(\wn)
 e^A_{kl}(\wn)\right),\nonumber \\
C_{2}(1)=
 \int\frac{\d^2 \wn}{4\pi}  \frac{\wn_a^i\wn_a^j\wn_b^k\wn_b^l}{1-(\wn\cdot\wn_a)^2}\left(
 \sum_{A=+,\times}e^A_{ij}(\wn)
 e^A_{kl}(\wn)\right).
 \label{cC0}
 \end{align} 
By using Eq. (\ref{wwww}), we find  that 
 \begin{align}
 C_{2}(0)&=\frac{1}{16\pi}\int \d^2 \wn\, \Big(1-\wn\cdot\wn_a\Big)^2,\nonumber \\
 C_{2}(1)&=\frac{1}{16\pi}\int \d^2 \wn\, \Big(1-(\wn\cdot\wn_a)^2\Big),
 \end{align}
 from which
 the relation 
 \begin{eqnarray}
 C_{2}(0)=2C_{2}(1)=\frac{1}{3}
 \end{eqnarray}
 follows, reproducing the coefficients in Eq. (\ref{HD3}).

\section{Proof that the integral of the HD correlation function over the celestial sphere vanishes}
In order to prove Eq. (\ref{HD2}), let us fix $\wn_b=(0,0,1)$ and let $\wn_a=(\sin\theta \cos\phi,\sin\theta\sin\phi,\cos\theta)$. Then from the definition of the two-point correlator in Eq. (\ref{cC}) we get
\begin{eqnarray}
 \label{HD01}
\int \d^2 \Omega_2 \, \Gamma_{ab}=\int \d^2 \wn_a \,\frac{\wn_a^i\wn_a^j}{1-\wn\cdot \wn_a}\times
\sum_{A=+,\times} \epsilon^A_{ij}(\wn)\times (\mbox{terms independent of $\wn_a$}). 
\end{eqnarray}
Now we want to evaluate the integral of the right-hand side of Eq. (\ref{HD01}), which due to symmetry should be written as
\begin{eqnarray}
\label{sss}
\int \d^2 \wn_a \,\frac{\wn_a^i\wn_a^j}{1-\wn\cdot \wn_a}=
\int \d^2 \wn_a \,\frac{\wn_a^i\wn_a^j}{1-s\, \wn\cdot \wn_a}
\Big|_{s=1}= A(s) n^i n^j+B(s) \delta^{ij}\Big|_{s=1},
\end{eqnarray}
where $A,B$ are constants. Since the above expression is contracted with the transverse and traceless polarization tensors 
$\epsilon^A_{ij}(\wn)$ we have that in general
\begin{eqnarray}
\Big(A(s) n^i n^j+B(s) \delta^{ij}\Big) \epsilon^A_{ij}(\wn)=0,
\end{eqnarray}
and therefore, 
\begin{eqnarray}
\label{innn}
\int \d^2 \wn_a \,\frac{\wn_a^i\wn_a^j}{1-\wn\cdot \wn_a}\sum_{A=+,\times} \epsilon^A_{ij}(\wn)=0
~~~~{\rm or}~~~~\int \d^2 \Omega_2 \, \Gamma_{ab}=0,
\end{eqnarray}
as claimed. Let us note that we have inserted the parameter $s\geq 1$ in Eq. (\ref{sss}) in order to avoid the pole at $\wn\cdot \wn_a=1$. We could equally use a $+i \epsilon$ shift in the denominator. Both ways lead to the vanishing of the integral in Eq. (\ref{innn}).

\newpage

\bibliographystyle{JHEP}
\bibliography{bib_v1}

\end{document}